# REAL TIME VOWEL TREMOLO DETECTION USING LOW LEVEL AUDIO DESCRIPTORS

*Mikhail Malt*
Ircam – MINT – Paris IV - Sorbonne
mikhail.malt@ircam.fr

*Marta Gentilucci*
Ircam - Harvard
marta@martagentilucci.com

## ABSTRACT

This paper resumes the results of a research conducted in a music production situation Therefore, it is more a final lab report, a prospective methodology then a scientific experience. The methodology we are presenting was developed as an answer to a musical problem raised by the Italian composer Marta Gentilucci. The problem was “how to extract a temporal structure from a vowel tremolo, on a tenuto (steady state) pitch.” The musical goal was to apply, in a compositional context the vowel tremolo time structure on a tenuto pitch chord, as a transposition control.In this context we decide to follow, to explore the potential of low-level MPEG7 audio descriptors to build event detection functions. One of the main problems using low-level audio descriptors in audio analysis is the redundancy of information among them. We describe an “ad hoc” interactive methodology, based on side effect use of dimensionality reduction by PCA, to choose a feature from a set of low-level audio descriptors, to be used to detect a vowel tremolo rhythm. This methodology is supposed to be interactive and easy enough to be used in a live creative context.

## 1. INTRODUCTION

This paper resumes the results of a research conducted in a music production situation Therefore, it is more a final lab report, a prospective methodology then a scientific experience.

The methodology we are presenting was developed as an answer to a musical problem raised by the Italian composer Marta Gentilucci. The problem was “how to extract a temporal structure from a vowel tremolo, on a tenuto (steady state) pitch.” The musical goal was to apply, in a compositional context the vowel tremolo time structure on a tenuto pitch chord, as a transposition control.

If the vibrato tracking has a wide scientific literature, we didn’t found any work related to the real-time detection of the vowel tremolo in a steady state pitch.

As a vowel tremolo can be seen as a formant tremolo, a logical path have could have been building a formant follower.

In the environment we worked (Max[1] 6.0), two tools were designed, and available, for this task: the first one was an old external developed by Serge Lemouton and Miller Puckette. After some preliminary tests it did not demonstrate stability and regularity in formant tracking. Because the performance conditions were complex, we decided to turn from that option (cf. 6.1).

The second available tool was the “lpc-toolkit”[2], a toolbox intended to support analysis and re-syntheses in real time. This tool could be used if we would take into consideration the variation of some filter coefficients.

However, we decide to follow another way, to explore the potential of low-level MPEG7 audio descriptors [1] to build event detection functions.

One of the main problems using low-level audio descriptors is the redundancy of information among them, even if the are very easy to calculate and not costly in CPU. Many of them are correlated and bring the same information (see [3] and [4]). The first step to work with audio descriptors is to reduce the dimensionality of the analytical data space and find what features are useful to describe the audio phenomena we are focusing on. Nowadays, exists a set of methodologies for feature selection, classification and dimensionality reduction, but, they are not yet implemented and adapted to be used in a creative process workflow. It does not exist yet a feature selection toolbox that is integrated in the main electroacoustic computer environments (Max, PD, SupperCollider, etc.). Even if we must agree that some tools exist, they are not well documented[3], and not integrated in a logical workflow as we conceived it. We needed a methodology that could help us in a “production process”, and adapted to a production time. With this paper, we would like to propose a simple methodology for dimensionality reduction and feature selection based on the PCA technic.

This paper is organized as follows: the next section present the musical context of composition, realization and production that gave birth to this research. Section 3 discusses some technological problems related with the real-time analytical implementation. Section 4 presents the main problems arising by working with audio features, especially the problem of information redundancy. Section 5 explains the reasons to choose the PCA technic. Section 6 details the interactive methodology proposed. Section 7 clarifies the use of the analytical data obtained. Section 8 discusses the results we obtained. We conclude the paper with some overall remarks on the methodology and about further developments.

---

[1] © cycling74.com.

[2] Max development by Mark Cartwright. http://www.markcartwright.com/projects/lpcToolkit/index.html

[3] Like the mnm.pca object from the FTM Max library. http://forumnet.ircam.fr/product/ftmco/

## 2. MUSICAL CONTEXT

The musical context which gave the first impulse to the research, was the composition of the piece *Da una crepa* (2011/12) for Soprano solo, Vocal Ensemble (five mixed voices: soprano, mezzo soprano, baritone and bass), Clarinet, Cello, Percussion and Live Electronics. The piece was premiered in June 2012 at the Centre Pompidou in Paris within the Manifeste Festival (IRCAM) by the *Cris des Paris* and the soloist of the Ensemble Intercontemporain.

In the past years, Marta Gentilucci has been working on some specific compositional questions regarding rhythmical structures: at the micro level, the relationship between the inner temporal structure of sound (very often vocal or vocal-like sounds) and the possibility to use that temporal structure as model for another static sounds (tenuto); at the macro level, the possibility to control and develop more complex rhythmical structures over a longer timespan of the composition.

For the piece *Da una crepa*, some specific technical/compositional boundaries were set: a rhythmical model was extracted from the analysis of micro-rhythmical structure of a singing voice which alternates in very legato a set of vowels couples (i-e-i-e or e-o-e-o, etc., see Figure 1); the data were used to reconstruct the same temporal structure, but additionally it was applied as transposition control to a tenuto sound of a singer.  It is to say, that if we record the voice which sings a vowels alternation, for example the sequence i-e-i-e, and then if we analyze the sound, the model resulting from the analysis could be applied on a simple tenuto sound so that it could control the amount of transposition variation maintaining the original time structure: shaping a tenuto sound by applying on it the rhythmical model of another. How to track the changes of the sonic content during the change of vowel and how/why the data were used as transposition control were central questions.

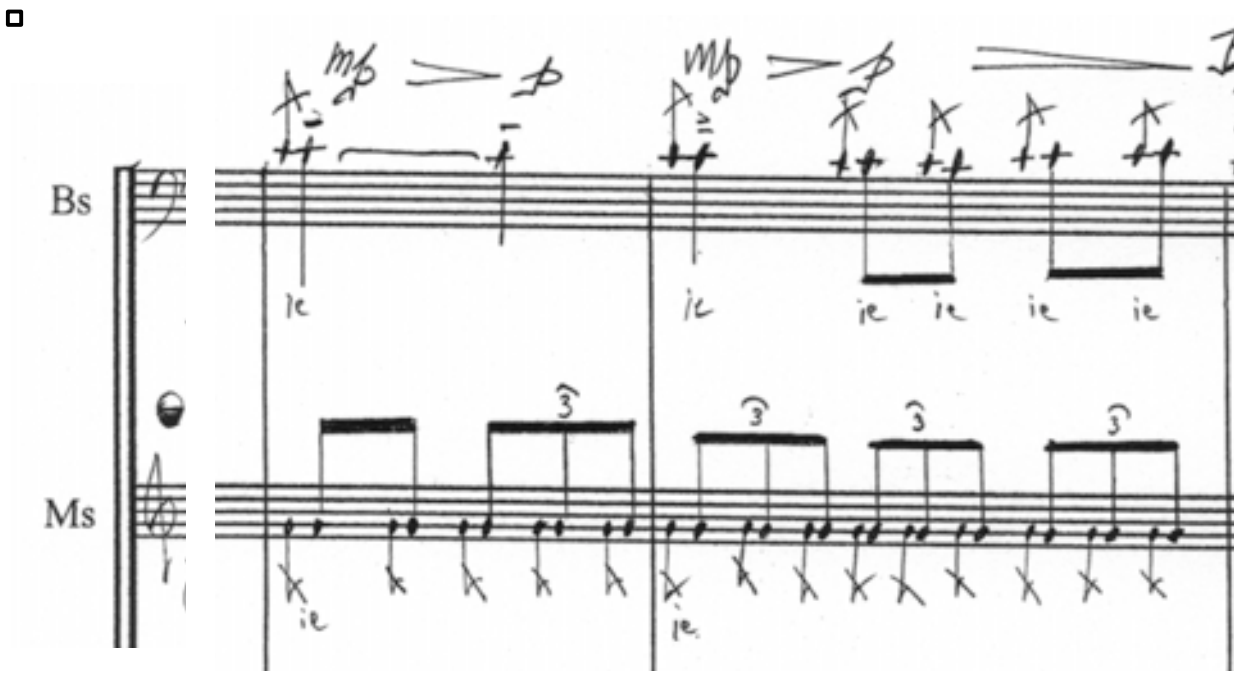

**Figure 1:** Vowel tremolo for Basse and Mezzo soprano, mesures 128-129, Da una crepa, Marta Gentilucci

## 3. TECHNOLOGICAL CONTEXT

The composer, Marta Gentilucci, needed to build a computer environment to manage the electronic sound in real-time. For this purpose we choose Max[1]. The "patch"[2] created for the performance contained a contained the following modules (sub-patches): two module of five independent spectral transposers, one module of five independent harmonizer, one module of spectral Delay, one module of resonance models, one module of multi-delay, one module of sixteen poly-rhythmical samples players, one module of Ircam SPAT[3] (five sources), one module of Ircam SPAT (eight sources), eighteen multi-channel sound files player. This engine list is just to show that the "concert patch" we needed was a complex environment and very demanding in CPU calculus. Then, the final technological implementation must take into account: the technologies ( treatments, synthesis models, etc.) which represent the composer 's musical needs, the electro-acoustical score and the efficiency of the implementation in terms of CPU. This technological point is very important because it was what leaded us to optimize our "analytical engine", designed to detect the vowel tremolo rhythm (with different vowels pairs, "i-o", "i-e, etc., and also with different voices, soprano, alto, bass, baritone, and tenor), and designed to share its computation needs with the other real-time audio processes.

## 4. WORKING WITH LOW LEVEL AUDIO DESCRIPTORS

One of the main problems working with audio descriptors is the information redundancy [Peeters & al. 2011, 2915], [Mitrovic & al. 2007]. For instance, spectral centroid, spectral slope and spectral roll-off, are highly correlated and it is very difficult to identify which descriptor, or descriptors set, will be more useful to describe a specific situation. In addition, there are very few research papers on their use as event detection functions in a real time performance context. Even if the number of toolboxes dedicated to audio features and audio descriptors computation increases, and even if we dispose of a good number of feature selection or dimensionality reduction algorithms, we found again a lack in methodologies dedicated to a real time performance context.

In a preliminary phase of our research we tried to find the best feature (or the best set of features) which described the vowel tremolo rhythm. We noticed that considering the voice type (soprano, mezzo soprano, baritone or bass), and according with the kind of different vowel tremolo (i-e, i-o) the "best feature" was not the same all the time.

From a technological point of view, a solution, could have been to choose a different feature for each voice and for each kind of tremolo but we rejected this possibility: this decision lead us to a less complex, and more stable computer environment.

[1] © Cycling74.com.

[2] A "patch" is a graphical program In Max's graphical programming environment.

[3] The SPAT is  "is a software suite for the spatialization of sound signals in real-time intended for musical creation, postproduction, and live performances" developed at Ircam. http://forumnet.ircam.fr/product/spat/?lang=en

At this point our problem was how to find "which" descriptor (feature) would better describe our phenomena.

In this context we developed a unified "ad hoc" methodology based on the dimensionality reduction technic called Principal component analysis (PCA) [cf. 5 and 6.3]. Thanks to this methodology, we identified the "main feature" supposed to be the best one to describe the rhythm (or the oscillations) of the vowel tremolo set.

## 5. WHY PCA?

The PCA is one of the oldest and best known [6, ix] techniques of multivariate analysis. Its main purpose is the dimensionality reduction of a large data set of inter-depended variables, by finding a new set of reference axis (eigenvectors), which describes the phenomena. Its main purpose is the dimensionality reduction of a large data set of interdependent variables. Its main purpose is the dimensionality reduction of a large data set of interdependent variables. Taking into account the amount of variance of each data set, a new set of reference axis (eigenvectors), which describes the studied phenomena, is obtained. Each axis, called principal component, is a linear combination of the original data. As shown, [4] "The quality of a feature could be measured by the amount of variation of its numerical values".

In our case we reduced the whole measure space (the set of all audio descriptors measured, cf. 6.2) to one feature. the one, which best represents our phenomena. This feature was the one that presented the higher modulo considering the first and the second principal components.

It is important to point out that the audio files used had mainly a tenuto pitch with a vowel tremolo. In the vowel tremolo sections (as page 18 from *Da una Crepa*, measures ) three voices are in a tenuto mode, while two of them do an eight-tone appoggiatura. Our hypothesis is that the main emergent phenomena (perceived and coming from the analysis) be this "vowel tremolo".

## 6. METHODOLOGY PROPOSED

The main analysis steps are:

High-quality audio recordings for each voice and for each vowel tremolo

Extraction of low-level MPEG-7 audio descriptors

PCA analysis,

Choice of the ten features, in each analysis, presenting the higher modulo between the first and second PCA (PCA1 and PCA2).

Features Ranking

Data visualization of the proposed analysis

### 6.1. Vowel tremolo recording

We recorded a set of audio files with the "vowel tremolo" in a performance situation. All the singers were simultaneously recorded using a "DPA 4066 Head set" cardioid microphone. Their position on stage was very similar to the actual concert situation. The result was a recording of the "vowel tremolo" but with the whole choir as background (Figure 2). The goal was to find the best feature in the "worst" situation.

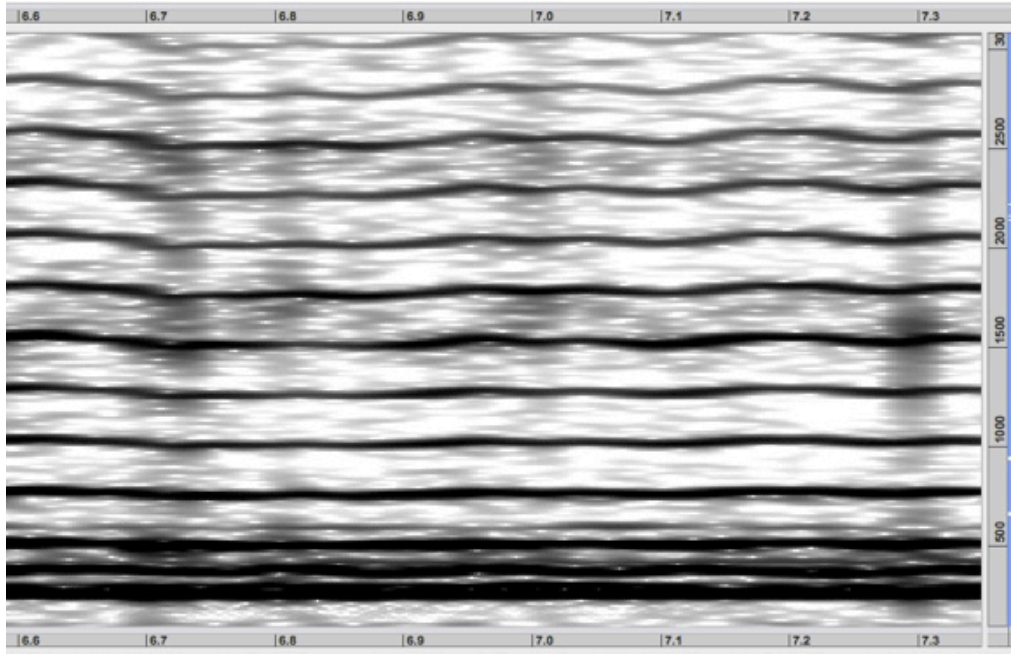

**Figure 2:** "i-e" tremolo by a tenor with the whole choir as background

### 6.2. Extraction of low-level MPEG-7 audio descriptors

For each audio file, a set of 42 low-level MPEG7 descriptors were computed. Actually, some descriptors (such as Tristimulus, SpectralFlatness and Spectral crest) have multiple components, which can be considered independent. Then, we could count 52 descriptors:[1]:

| | |
|---|---|
| HarmonicSpectralRolloff | HarmonicSpectralSlope |
| HarmonicSpectralDecrease | HarmonicSpectralVariation |
| HarmonicSpectralKurtosis | HarmonicSpectralSkewness |
| HarmonicSpectralSpread | HarmonicSpectralCentroid |
| HarmonicSpectralDeviation | HarmonicTristimulus |
| HarmonicOddToEvenRatio Inharmonicity | Noisiness |
| NoiseEnergy HarmonicEnergy | FundamentalFrequency |
| SpectralFlatness SpectralCrest | PerceptualSpectralRolloff |
| PerceptualSpectralSlope | PerceptualSpectralDecrease |
| PerceptualSpectralVariation | PerceptualSpectralKurtosis |
| PerceptualSpectralSkewness | PerceptualSpectralSpread |
| PerceptualSpectralCentroid | PerceptualTristimulus |
| PerceptualOddToEvenRatio | PerceptualSpectralDeviation |
| Sharpness Spread Loudness SpectralRolloff | SpectralSlope |
| SpectralDecrease SpectralVariation | SpectralKurtosis |
| SpectralSkewness SpectralSpread | SpectralCentroid |
| TotalEnergy SignalZeroCrossingRate | |

**Table 1:** Low level MPEG-7 audio descriptors used

The input representation chosen for the feature calculation was the default one, linear magnitude and linear frequency. The feature calculation returned a 52 columns matrix $M_i$, where each column, was the temporal evolution of a feature. "i" is the index representing a specific audio file, analysed.

Each matrix $M_i$ had theirs columns smoothed by a low pass filter, returning a matrix $M_i$s.

[1] This computation was done with Ircamdescriptor~, a Max implementation of the "Timbre Toolbox" [3]. For a precise description of each feature, please refer to [1]. We are also using the same denomination as in [1].

### 6.3. PCA analysis

On each data matrix, $M_i$s, a PCA (principal component analysis) was computed[1], returning the $M_i$s.eigen matrix, the eigenvector matrix associated to audio file "i". From this matrix we keept the first two columns, representing the first and the second principal components (pca1 & pca2), $M_i$s.eigen1 and $M_i$s.eigen2. According to the analysis, these two components explained between 70 and 79% of the cumulated variance, with the first component explaining between 55 and 69% of the total variance. It seemed sufficient to keep just these two first principal components.

### 6.4. The bests features

For each feature represented as an index row in $M_i$s.eigen1 and $M_i$s.eigen2, the modulo from its first and second principal components is computed. It returns a new matrix $M_i s.\text{mod}[k,j]$. The first column "k" contains the descriptors' names, and the second column the modulo (or the Euclidian distance) "j", in a two dimensional space, where the orthogonal axis are pca1 and pca2:

$$M_i s.\text{mod}[k,j] = \sqrt{M_i s.eigen[k,1]^2 + M_i s.eigen[k,2]^2}\ .$$

We can associate each "j" value from $M_i s.\text{mod}[k,j]$ to the expressiveness of the "k" feature.

Figure 3 shows the two-dimensional plot for the first nine $M_i s.\text{mod}[k,j]$ for an "i-e" vowel tremolo sung by a tenor. In the same figure, we can observe that some features are strongly correlated. For example, "Perceptual tristimus2" and "Harmonic tristimus1" are correlated, "Total energy" and "Harmonic energy" are so correlated that they are superposed. On the contrary, we can see that Perceptual tristimus2 and Perceptual tristimus1 are in an inverse relationship and in an orthogonal relationship with Spectral Flatness3.

### 6.5. Features Ranking

From the obtained data, we sorted each one of $M_i s.\text{mod}[k,j]$, creating $M_i s.\text{mod}.sort[k,j]$, where the index ($index_i(k)$) of each feature represents the rank of the feature according to its "j" ($M_i s.\text{mod}[k,j]$) value. The matrix $index_i(k) = 1$ represents the feature "k" in the audio file analysis "i" having the greater "j" value. A new matrix $D_i(k, j, w_i(k))$ was created, containing, the first ten descriptors from $M_i s.\text{mod}.sort[k,j]$, where

$w_i(k) = 11 - index_i(k)$, is the grade of the "k" feature in the audio file analysis "i".

[1] PCA calculus was done in the R environment with < prcomp()> command, with zero centered variables and rescaled variables with unit variance.

For each "k" feature a mean weight $\overline{w}(k) = \frac{1}{n}\sum_{i=1}^{n} w_i(k)$ was computed, where $\overline{w}(k)$ represents, the contribution of the feature "k" to explain the studied phenomena.

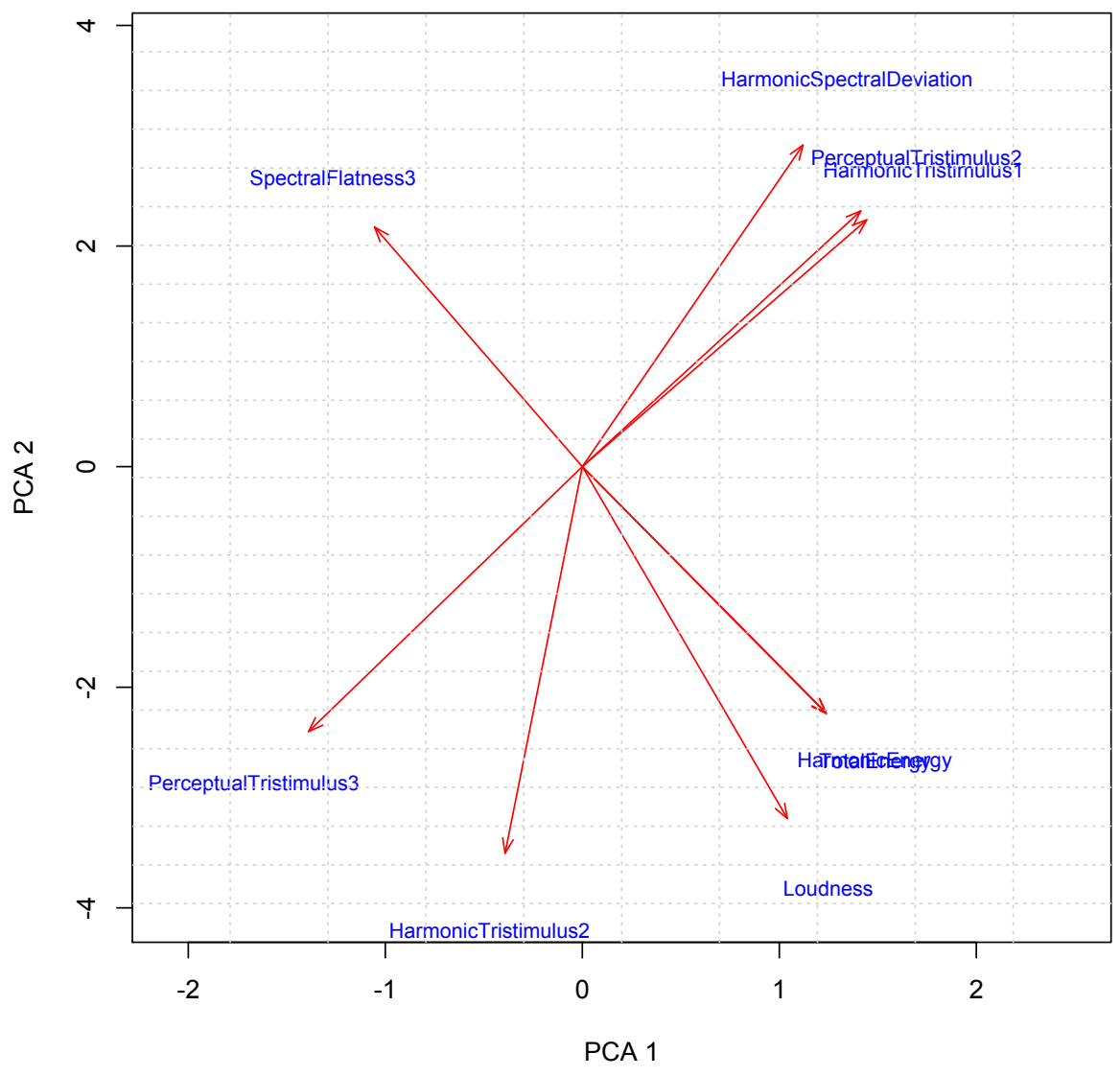

**Figure 3:** Two dimensional plot for Tenor "ie" vowel tremolo

The final result could be seen in Table **2**.

| « k » feature | $\overline{w}(k)$ |
|---|---|
| harmonictristimulus2 | 7.2 |
| loudness | 6.0 |
| totalenergy | 5.6 |
| harmonicenergy | 5.2 |
| harmonicspectraldeviation | 4.2 |
| perceptualtristimulus3 | 3.8 |
| harmonicspectraldecrease | 3.2 |
| harmonicoddtoevenratio | 3.2 |
| perceptualtristimulus2 | 3.0 |
| harmonicspectralskewness | 2.8 |
| harmonictristimulus1 | 2.2 |
| spread | 2.0 |
| spectralkurtosis | 2.0 |
| fundamentalfrequency | 1.8 |
| harmonicspectralkurtosis | 1.0 |
| perceptualoddtoevenratio | 1.0 |
| spectralflatness3 | 0.4 |
| spectralcrest3 | 0.4 |

Table 2**: Final descriptors rank**

In our case the best candidate was the harmonictristimulus2, the second second tristimulus. The second tristimulus is the ratio between the energy of the second, the third and the fourth harmonics, on the total harmonic energy in a signal [1],

$$T2 = \frac{a(2) + a(3) + a(4)}{\sum_h a(h)}\ .$$

Even if we presented all the scores computed, according with our electroacoustic real-time music performance practice, we preferred just to keep descriptors more related with the spectral shape than the overall sound energy. The descriptors related to the signal energy (harmonicEnergy, loudness and totalEnergy) where suppressed, in our final ranking.

In a real time performance context the background sound is a parameter difficult to control (see [2]). Even if the singers were using very good headset cardioid microphones (dpa 4066), the sound capture is was not clean because it took also the background sound, the electronic sound, etc. The precise level control is difficult in (real) real-time electronic performances.

So, discarding the signal energy related features, the score reached by the second tristimulus (7.2/10) is almost 70% bigger than the next candidate, the harmonicspectraldeviation that reached a score of 4.2/10.

### 6.6. Data visualization

The last step in our methodology was to verify experimentally if this analysis could return a clean data signal, and to check if the variation register has enough precision to build an event detection function.

The score obtained by the second tristimulus does not mean that it was the best feature in each audio file analysis, but it was the feature being able to have the best performance in these different situations including five different voices and two types of vowel tremoli. As we can see in Figure 4, the feature contains noise: it usually dues to audio interferences between the singers and the amplitude change. In the final implementation of this analysis (second tristimulus) in real time, second tristimulus was triggered after the stabilization phase of the vowel tremolo, and stopped before its end, in order to avoid interferences and to keep the analysis with the best ratio between the singer audio level and the background audio level.

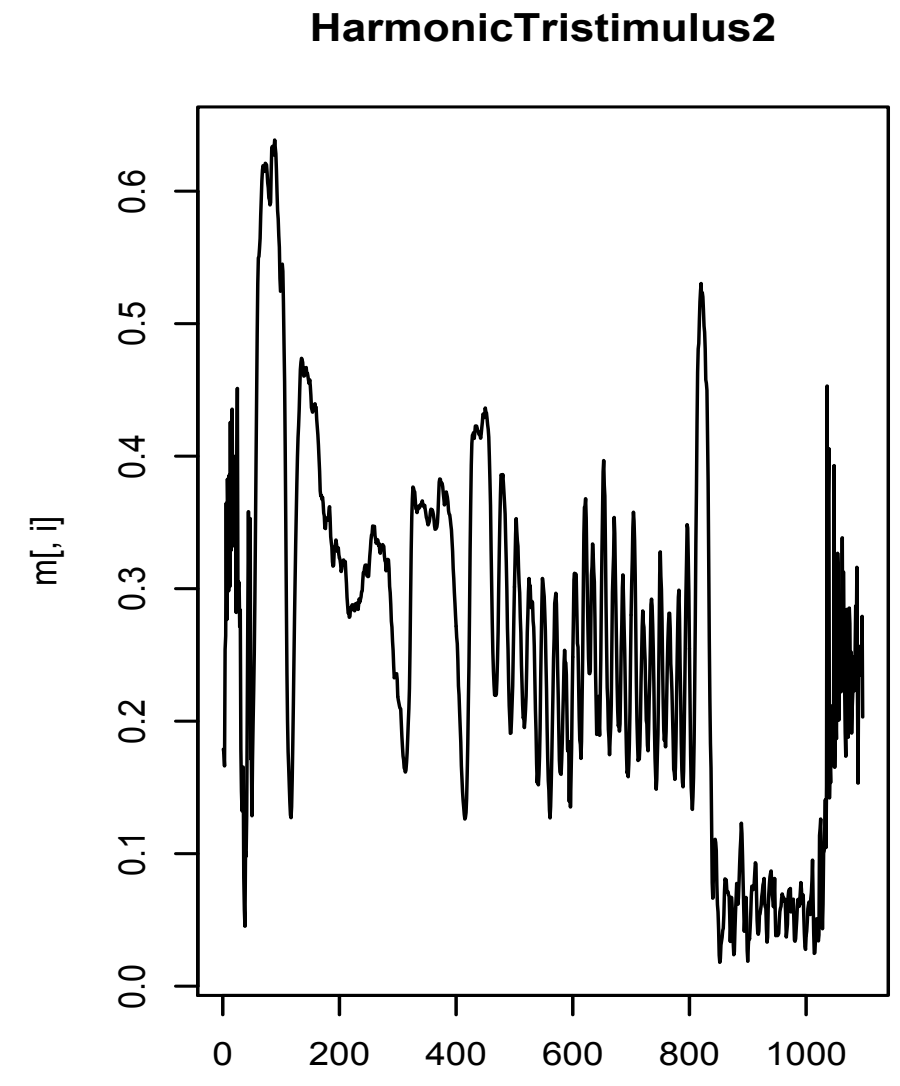

**Figure 4:** Second harmonic tristimulus for an "i-e" vowel tremolo by a tenor

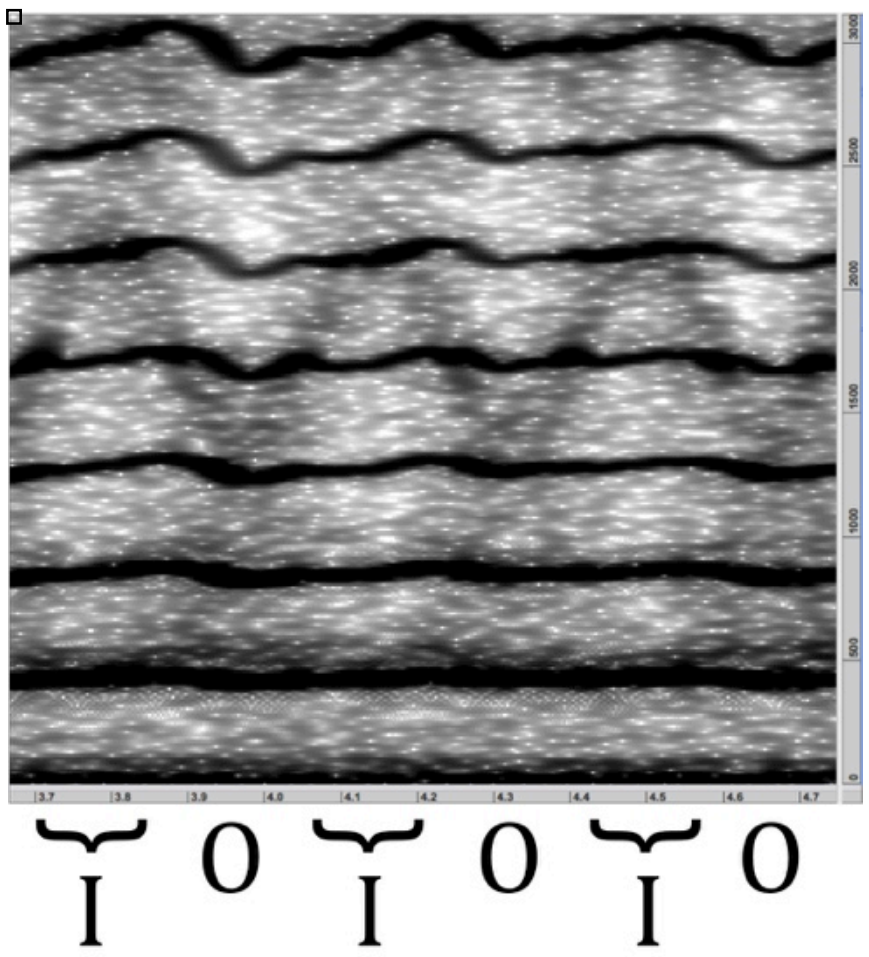

**Figure 5:** Soprano "io" vowel tremolo around A4

## 7. RYTHMIC DETECTION AND TRANSPOSITION

After the second tristimuls analysis, we built a standard event detection function by standard derivative calculation and peak-picking techniques as is found in [2] and [7]. The data was also kept in data buffers (<coll> objects in Max) to be reused as transposition control. Each analysis was zero centered and normalized, in order to be able to control the register transposition oscillation wanted.

## 8. DISCUSSION

As we can see in Figure 5 and Figure 6, the vowel tremolo used, "i-o" and "i-e", causes a visible energy variation between the second and the sixth harmonics. So, we can understand the results as being an indirect measure of the energy variation in the zone of the second, third and fourth harmonics, caused by the vowel tremolo.

An other vowel tremolo side effect we can observe, is the fact that the tremolo are between two vowels with different sound colors. The bright "i" and the darks "e" and "o". In the two cases, this sound color change seems to induce the singers in a small frequency oscillations, where the "e" and "o" are lower than the "i". Just to remember, the singer was supposed to make their vowel tremolo on a tenuto pitch.

In our case, we tried to reduce the space dimensionality of our analysis to a single feature. It was possible because we studied audio examples with a very clear behavior, a steady state pitch with a vowel tremolo. Our main hypothesis was that the main phenomena were the vowel tremolo, and that the PCA analysis will return the feature explaining, or expressing this movement.

But in other cases, other audio, or sound textures with more complex behavior, one single feature could not be sufficient. In this case the study of the redundancy and the features orthogonality, could be easily done using a n-dimensional study of the PCA results, like in Figure 3.

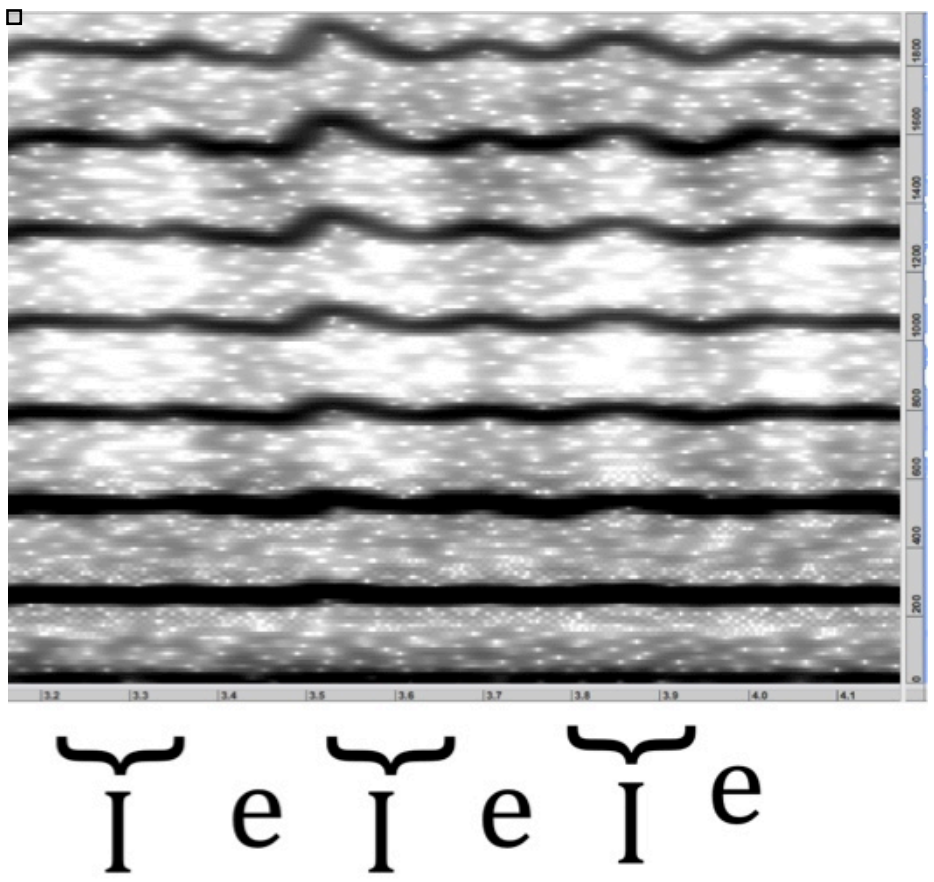

**Figure 6:** Bass "ie" vowel tremolo around C4

## 9. CONCLUSIONS

We have described an "ad hoc" interactive methodology, based on a side effect use of dimensionality reduction by PCA, to choose a feature from a set of low-level audio descriptors, to be used to detect a vowel tremolo rhythm. This methodology is supposed to be interactive and easy enough to be used in a live creative context.

The proposed PCA procedure is a very easy and ongoing practice that could be adapted and used in the situation we explained. Even if we focus on a specific feature, it seems clear that this methodology could also be applied to other situations to find the best, or the bests features.

Another point that emerges from this experience, is the fact that event if this technique was mainly intended to data dimensionality reduction, it also could be used to identify redundancy in data analysis, identifying and choosing an orthogonal set of features, as we showed [cf. item 6.4 and Figure 3].

The main difficulty we found in the process was the workflow: it asked to us to go from one software to another. We needed to build three different environments: a first environment (using Max) for the descriptors analysis, visualization, and data export, a second programming environment (in R) to proceed to the PCA analysis, visualize the data, sort and rank it, and a third environment (in Max) to proceed to test to check the data proposed by the PCA. Now that we could verify the utility of this method, our next step will be to build a Max toolbox to be used by a single musician with a single computer, working with audio features to build "event detection functions".